\documentclass[usenatbib]{mnras}

\usepackage{newtxtext,newtxmath}

\usepackage[T1]{fontenc}
\usepackage{ae,aecompl}
\usepackage{tikz}
\usepackage{pgfplots}
\usepackage{diagbox}
\usepackage{multirow}

\usepackage{amsmath}	
\usepackage{amssymb}	
\DeclareMathOperator*{\argminA}{arg\,min} 
\usepackage[utf8]{inputenc}
\usepackage{hyperref}
\usepackage{color}
\usepackage{ulem}

\pgfplotsset{compat=1.14}

\title[Estimation of atmospheric turbulence parameters]{Estimation of atmospheric turbulence parameters from Shack-Hartmann wavefront sensor measurements}

\author[P. P. Andrade et al.]
{Paulo P. Andrade,$^{1,2}$\thanks{E-mail: ppandrad@fe.up.pt, pgarcia@fe.up.pt}
Paulo J. V. Garcia,$^{1,2,4}$
Carlos M. Correia,$^{3}$
Johann Kolb$^{4}$
and Maria In\^{e}s Carvalho$^{1,5}$
\\
$^{1}$Faculdade de Engenharia da University of Porto, Rua Dr. Roberto Frias s/n, 4200-465 Porto, Portugal\\
$^{2}$CENTRA – Centro de Astrofísica e Gravitação, IST, Universidade de Lisboa, 1049-001 Lisboa, Portugal\\
$^{3}$Aix Marseille Univ, CNRS, CNES, LAM, Marseille, France\\
$^{4}$European Southern Observatory, Alonso de Córdova 3107, Vitacura, Región Metropolitana, Chile\\
$^{5}$INESC TEC, Rua Dr. Roberto Frias, 4200-465 Porto, Portugal
}

\date{Accepted XXX. Received YYY; in original form ZZZ}

\pubyear{2018}

\begin{document}
\label{firstpage}
\pagerange{\pageref{firstpage}--\pageref{lastpage}}
\maketitle

\begin{abstract} 

The estimation of atmospheric turbulence parameters is of relevance for: a) site evaluation \& characterisation; b) prediction of the point spread function; c) live assessment of error budgets and optimisation of adaptive optics performance; d) optimisation of fringe trackers for long baseline optical interferometry.
The ubiquitous deployment of Shack-Hartmann wavefront sensors in large telescopes makes them central for atmospheric turbulence parameter estimation via adaptive optics telemetry. Several methods for the estimation of the Fried parameter and outer scale have been developed, most of which are based on the fitting of Zernike polynomial coefficients variances reconstructed from the telemetry. The non-orthogonality of Zernike polynomial derivatives introduces modal cross coupling, which affects the variances. Furthermore, the finite resolution of the sensor introduces aliasing. In this article the impact of these effects on atmospheric turbulence parameter estimation is addressed with simulations. It is found that cross coupling is the dominant bias. An iterative algorithm to overcome it is presented.
Simulations are conducted for typical ranges of the outer scale (4 to 32\,m), Fried parameter (10\,cm) and noise in the variances (signal-to-noise ratio of 10 and above). It is found that, using the algorithm, both parameters are recovered with sub-percent accuracy.

\end{abstract}




\section{Introduction}

The atmosphere is a refractive fluid in turbulent motion. Light waves, after propagating through it, exhibit random phase perturbations and, if over long distances, random amplitude variations. Atmospheric turbulence is generally modelled as a stochastic process with structure functions given by an empirical turbulence model \citep{Hardy1998, Roddier1999}. In most spatial scales turbulence roughly obeys the Kolmogorov model. However, because atmospheric turbulence cascades from eddies of a effective maximum size, an outer scale for the turbulence has to be introduced, where turbulent effects saturate. Several options exist \citep[e.g.][]{Voitsekhovich1995} for  adapting the Kolmogorov model by including an outer scale in the structure function, a widely used one being the von Kármán model.  Using the index of refraction structure function, the  power spectrum of the optical field phase fluctuations can be computed \citep[e.g.][]{Conan2000}, being given by
\begin{equation}\label{eq:W_phi}
W_\phi(f)= \alpha \left(f^2+\frac{1}{\mathcal{L}_0^2}\right)^{-11/6} r_0^{-5/3} ,
\end{equation}
where $f$ is the spatial frequency,  $\alpha$ is a constant\footnote{
$\alpha=\frac{\Gamma^2(11/6)}{2\pi^{11/3}} \left[\frac{24}{5}\Gamma\left(\frac{6}{5}\right)\right]^{5/6} \simeq 2.2896\times 10^{-2}$.}, $\mathcal{L}_0$ is the outer scale\footnote{This parameter of the von Kármán model is not to be confounded with the spatial scale of the large turbulent eddies. In this work the wavefront coherence outer scale will referred as simply outer scale \citep[cf.][for a discussion of the outer scale definitions]{Ziad2016}.} and $r_0$ is the Fried parameter
\begin{equation}
r_0(\lambda) = \beta \left(\frac{2\pi}{\lambda}\right)^{-6/5}
\left[\int_0^\infty dz C_\mathrm{n}^2(z)\right]^{-3/5},
\end{equation}
where  $\beta$ is a constant\footnote{$\beta= \left[\frac{5\Gamma(2/3)}{\sqrt{\pi}\Gamma(1/6)}\right]^{3/5} \sqrt{\frac{24}{5}\Gamma(\frac{6}{5})}\simeq 1.6748$.}, $\lambda$ is the optical beam wavelength and $C_\mathrm{n}^2(z)$ the refraction index structure constant in function of height $z$. In this framework the optical phase fluctuations are characterised with only two turbulence parameters: the outer scale $\mathcal{L}_0$ and Fried's parameter $r_0$. In particular, it explains why the long exposure point spread function of the largest optical-infrared telescopes is limited by the atmosphere to that of an equivalent telescope of a few tens of centimetres \citep[cf. the image quality of seeing-limited surveys][]{Castellano2010,Libralato2014} and is a function of both the outer scale and Fried's parameter \citep[e.g. study of][]{Martinez2010}.

The above framework is approximate. The turbulence strength, as given by $C_\mathrm{n}^2(z)$, is stratified in layers with different speeds \citep[e.g.][]{Osborn2017,Osborn2018}. These layers can have different outer scales $\mathcal{L}_0(z)$ \citep[e.g.][]{Maire2007,Guesalaga2017}. 
Atmospheric turbulence is non-stationary and the above parameters are found to vary in time
\citep[e.g.][]{DaliAli2010,Floyd2010,Maire2006,Ziad2004}. The surface layer of atmospheric turbulence appears to not follow Kolmogorov turbulence \citep[e.g.][]{Lombardi2010} as well as the lower altitude dome-related related turbulence \citep[e.g.][]{Guesalaga2014,Helin2018}. Deviations from the Kolmogorov refractive index structure function or phase power spectrum power law coefficients have been referred to by several authors \citep{Dayton1992,Nicholls1995,Goodwin2016}, 
but it is not clear if these are manifestations of the outer scale or data filtering effects. 
In general, the framework is roughly consistent with measurements 
\citep[e.g.][]{Dayton1992,Rao1999,Tokovinin2007} and found to be a working solution and of standard use in high angular resolution (adaptive optics or interferometry) simulations, system design and data-reduction. 
 
Optical characterisation of atmospheric turbulence, i.e. the determination of the atmospheric turbulence
parameters, is relevant in several instances: 
a) site evaluation \& characterisation; 
b) optimisation of AO systems, whose performance can only be assessed when
the turbulence conditions under which they are operating are known; 
c) making predictions of point spread functions in a variety of conditions 
(with or without AO); 
d) optimisation of fringe trackers for optical interferometry \citep[e.g.][]{Conan2000b, Choquet2014}.

Several methods are available for the characterisation of the turbulence 
\citep[cf. review of][on Balloons, DIMM, MASS, SLODAR, SCIDAR, etc.]{Lombardi2014}. 
In this article, the estimation of turbulence parameters from the telemetry of an AO Shack-Hartmann wavefront sensor (SH-WFS) is considered. Their  ubiquity in large telescopes  makes the development of  methods for AO telemetry attractive as they have the advantage of making use of existing  infrastructure. On top of that, using AO telemetry is more appropriate to circumvent inconsistencies related to temporal synchronism, angular co-alignment and identical turbulence path (e.g. including dome seeing) of the observations. These are issues that affect the former methods. The use of AO telemetry for atmospheric turbulence sensing is not new \citep[e.g.][]{Schoeck2003, Fusco2004, Jolissaint2018} for single-sensor telemetry and \citep{Guesalaga2017, Ono2017} for multiple sensor telemetry. 

The common approach to estimate turbulence parameters from a single SH-WFS data uses the projection on Zernike polynomials of the wavefront phases reconstructed from the telemetry data  -- wavefront phase gradient measurements and deformable mirror commands. The turbulence parameters are retrieved by fitting a theoretical Zernike representation of the turbulence spectrum to the variances of the Zernike coefficients. Noise, aliasing and modal cross coupling \citep{Herrmann1981} are effects introduced by the SH-WFS which degrade the quality of the wavefront reconstructions and limit the accuracy of the method. Aliasing results from the finite spatial resolution of the SH-WFS when sampling phase gradients of turbulence-induced distorted phases which have non-bandlimited spatial frequency content. 
Cross coupling has its origin in the non-orthogonality of the first-order derivatives of the Zernike polynomials over the circular pupil. In the phase reconstruction process, which is based on a least squares fit of these derivatives to the phase-gradients, the non-orthogonality gives rise to a matrix of
normal equations which is not diagonal, and thus, originates modal cross coupling. Since the variances of the reconstructed Zernike coefficients incorporate contributions from noise, modal cross coupling and aliasing, the turbulence parameters retrieved from the fit of the theoretical variances are biased.

In this article, the effects of aliasing and modal cross coupling on the variances of the reconstructed Zernike coefficients are studied with simulations. An algorithm to remove the  impact of cross coupling in the estimation of the turbulence parameters is proposed. It relies on the correction of the modal variances for cross coupling before the fit and, since its calculation requires the knowledge of the turbulence parameters,it has an iterative nature.
In the next section the methods used for turbulence parameter estimation are presented. In Section~3 we start by showing that aliasing has a negligible impact but cross coupling significantly biases the measurements. It is then shown than the proposed iterative algorithm converges to the simulation parameters in 3 iterations, with sub-percent accuracy. In Section~4 we conclude and discuss possible developments.

\section{Methods}

\subsection{Shack-Hartmann model and phase reconstruction}

In a modal representation, the wavefront phase is given by a series expansion in terms of basis functions. Zernike polynomials, orthogonal within a circle of unit radius, are a common choice for basis functions due to the similarity of the low order polynomials with familiar optical aberrations. Using Zernike polynomials with Noll's numbering and normalisation \citep{Noll1976}, the phase is given by
\begin{equation}
\phi(R\rho,\theta) = \sum_{i=2}^M a_i Z_i(\rho,\theta)\,,
\end{equation}
where $Z_i$ is the Zernike polynomial of order $i$, $R$ is the telescope radius, $\rho$ and $\theta$ the  polar coordinates in the unit circle.  The summation starts at $i = 2$ since the first term (piston mode) is of no concern for a SH-WFS, and ends at $M$, a sufficiently large number for  the approximation to the infinite sum. 

If the phase at the telescope pupil is represented by  a column vector of Zernike coefficients, $\mathbf{a} = [a_2,\cdots,a_M ]^t$,  its measurement by a SH-WFS with $N$ sub-apertures is modelled by
\begin{equation}\label{directModel}
\mathbf{s} = \mathbf{G}\mathbf{a} + \mathbf{w}\,,
\end{equation}
where $\mathbf{s} = [s_1,\cdots,s_{2N}]^t$ is a column vector of $2N$ slopes, $\mathbf{G}$ is a $2N\times(M - 1)$ matrix that describes the wavefront sensor response to the Zernike modes and $\mathbf{w} = [w_1,\cdots,w_{2N}]^t$ is a column vector representing the measurement noise.

The estimation of the phase from the slope measurements requires an inversion of the direct problem expressed by equation~(\ref{directModel}).

If $J$ is the order of the highest mode to be estimated, let $\mathbf{a}$ be represented by the concatenation $\mathbf{a}=[\mathbf{a}_\varparallel \mathbf{a}_\perp]^t$ where $\mathbf{a}_\varparallel=[a_2,\cdots, a_J]^t$ and $\mathbf{a}_\perp = [a_{J+1},\cdots,a_M]^t$. Similarly, let $\mathbf{G}$ be given by
$\mathbf{G} = [\mathbf{H}_\varparallel \mathbf{H}_\perp]$ where $\mathbf{H}_\varparallel$ is the matrix containing the first $J-1$ columns of $\mathbf{G}$, and $\mathbf{H}_\perp$ the matrix formed by the remaining columns. Equation~(\ref{directModel}) becomes
\begin{align}\label{direct_model_a}
	\mathbf{s}  & =  \begin{bmatrix}
							\mathbf{H}_\varparallel & \mathbf{H}_\perp \\
						\end{bmatrix}
						\begin{bmatrix}
							\mathbf{a}_\varparallel \\ 
    						\mathbf{a}_\perp
						\end{bmatrix}
						+\mathbf{w} \nonumber \\
				& = \mathbf{H}_\varparallel \mathbf{a}_\varparallel + 
                    \mathbf{H}_\perp \mathbf{a}_\perp +  \mathbf{w}\,.
\end{align}
The least squares solution of equation~(\ref{direct_model_a}) is given by 
\begin{equation}
\mathbf{b} = \mathbf{H}^+ \mathbf{s}\,, 
\end{equation}
where $\mathbf{b}=[b_2,\cdots,b_J]^t$ is the column vector with the estimates of the first $J-1$ Zernike coefficients $\mathbf{a}_\varparallel$ and $\mathbf{H}^+$, the reconstruction matrix, is the generalised inverse of matrix $\mathbf{H} _\varparallel$, which is given by
\begin{equation}
\mathbf{H}^+ = 
(\mathbf{H}_\varparallel^t \mathbf{H}_\varparallel)^{-1}\mathbf{H}_\varparallel^t\,. 
\end{equation}
The relation between true (atmospheric, $\mathbf{a}_\varparallel$) and estimated ($\mathbf{b}$) Zernike coefficients, becomes \citep{Dai1996}
\begin{equation}\label{reconstructed_modes}
 \mathbf{b} = \mathbf{a}_\varparallel  
+ \mathbf{H}^+ \mathbf{H}_\perp \mathbf{a}_\perp +
\mathbf{H}^+\mathbf{w} \,. 
\end{equation}
The second term of equation~(\ref{reconstructed_modes}) results from the truncation of matrix $\mathbf{G}$ and shows how the estimated lower modes are affected by the higher ones not present in the reconstruction matrix. This  cross coupling effect is caused by the lack of orthogonality between columns in matrix $\mathbf{G}$ which are, essentially, the gradients of the Zernike polynomials \citep[non orthogonal over the circular aperture,][]{Herrmann1981,Janssen2014}.

\subsection{Zernike coefficient variances}

The covariance matrix for the estimated Zernike coefficients is 
\begin{equation}\label{zernikeCovariances}
\langle \mathbf{b}\mathbf{b}^t \rangle = 
\langle \mathbf{a}_\varparallel\mathbf{a}_\varparallel^t\rangle +
\mathbf{C}\langle \mathbf{a}_\perp\mathbf{a}_\perp^t\rangle \mathbf{C}^t +
2\mathbf{C}\langle\mathbf{a}_\perp\mathbf{a}_\varparallel^t \rangle + 
\mathbf{H}^+ \langle \mathbf{w} \mathbf{w}^t \rangle(\mathbf{H}^+)^t\,,
\end{equation}
where $\langle . \rangle$  represents the expectation value and $\mathbf{C}=\mathbf{H}^+ \mathbf{H}_\perp$ is a cross-talk matrix, as defined by \citet{Dai1996}. It is assumed that there is no correlation between  measurement noise and the Zernike coefficients ($\langle \mathbf{w}\mathbf{a}_\varparallel^t \rangle = \mathbf{0}$ and $\langle \mathbf{w}\mathbf{a}_\perp^t \rangle = \mathbf{0}$).

The diagonal elements of equation~(\ref{zernikeCovariances}) give the relation between reconstructed and atmospheric Zernike coefficient variances.  Representing $h_{ij}^+$ and $c_{ij}$ as the matrix elements of, respectively,  the reconstruction and the cross-talk matrices
\begin{equation}\label{zernike_variances}
	 \langle b_i^2 \rangle =  \langle a_{\varparallel i}^2 \rangle +
    \sigma_{\mathrm{cc},i}^2  + \sigma_{\mathrm{n},i}^2\,,
\end{equation}
where,
\begin{equation}\label{sigmaa}
          \sigma_{\mathrm{cc},i}^2 = \sum_{j = J + 1}^M  \sum_{j' = J + 1}^M  
              c_{ij} \langle a_{\perp j} a_{\perp j'} \rangle c_{j'i}^t + 
              2 \sum_{j = J + 1}^M c_{ij} \langle a_{\varparallel i} a_{\perp j} \rangle \,,
\end{equation}
is the contribution to the modal variances\footnote{In the article modal variances and Zernike coefficient variances are used interchangeably.} associated with cross coupling  \citep[which can be computed with the expressions of][and are functions of $r_0$ and $\mathcal{L}_0$]{Takato1995,Conan2000}.
The contribution to the modal variances associated with noise, given by
\begin{equation}\label{sigmaw}
          \sigma_{\mathrm{n},i}^2 = \sum_{j = 1}^{2N}  \sum_{j' = 1}^{2N}  
              h_{ij}^+ \langle w_j w_{j'} \rangle (h^+)_{j'i}^t \,, 
\end{equation}
can be simplified by assuming that the noise affecting slope measurements is uncorrelated and has equal  variance $\sigma_0^2$ \citep{Southwell1980}
\begin{equation}\label{noiseterm}
\sigma_{\mathrm{n},i}^2  \approx 
\sigma_{0}^2 [\mathbf{H}_\varparallel^t \mathbf{H}_\varparallel]^{-1}_{ii}\,.
\end{equation}

\subsection{Algorithm for turbulence and noise parameters estimation}\label{algorithm}

The turbulence parameters are retrieved from a fit to the variances of the reconstructed Zernike coefficients. The method is reminiscent  of that in \cite{Fusco2004} but generalises it to mitigate biases stemming from modal cross coupling. Noise and cross coupling in the reconstructed modal variances are dealt in two  different ways. Noise is included in the modelling of the modal variances by adding its associated variances, equation~(\ref{noiseterm}), to the theoretical von Kármán Zernike coefficient variances. The fitting function, $f(\mathbf{p})$, is, thus,
\begin{equation}\label{fitfunction}
f(\mathbf{p})=\left(\langle a_{\varparallel i}^2 \rangle_{\mathrm{vK}} + \sigma_{\mathrm{n},i}^2 \right)(\mathbf{p})\,, 
\end{equation}
where $\mathbf{p}=[r_0,\mathcal{L}_0,\sigma_0^2]$ is a vector including the turbulence parameters and a noise parameter. A similar modelling of cross coupling by inclusion of the respective term in the fitting function would be unpractical since it would require the manipulation of large matrices at each iteration of the least squares algorithm. Instead, the disturbances associated with cross coupling, given by equation~(\ref{sigmaa}), are computed and removed  from the modal variances. In the same spirit as the procedure used in \cite{Veran1997}, since these calculations require the knowledge of the turbulence parameters, the same for the estimation of which  they are being made, the following iterative approach is adopted
\begin{align}\label{lsqFit1}
\hat{\mathbf{p}}^k =  \argminA_{\mathbf{p}} 
\sum_{i=5}^{J(r)}  \Bigg \{ &
\log\left[ (\langle a_{\varparallel i}^2 \rangle_{\mathrm{vK}}  +
         \sigma^2_{\mathrm{n},i})(\mathbf{p}) \right] -\\ \nonumber
& \log\left[ \langle b_i^2 \rangle -\sigma_{\mathrm{cc},i}^2(\hat{\mathbf{p}}^{k-1}) \right]
\Bigg \}^2\,,\,\,\, k = 1,\ldots
\end{align}
where  $\hat{\mathbf{p}}^k$ is the vector of parameter estimates at iteration $k$, $J(r)$ is the number of Noll modes up to radial order $r$\footnote{Here the radial order is denoted by $r$ for clarity, in contrast with other works where it is denoted by $n$.}, and
\begin{equation}\label{lsqFit2}
\hat{\mathbf{p}}^0 =  \argminA_{\mathbf{p}} 
\sum_{i=5}^{J(r)} \{ 
\log\left[(\langle a_{\varparallel i}^2 \rangle_{\mathrm{vK}}+
\sigma^2_{\mathrm{n},i})(\mathbf{p})\right]-\log\left[\langle b_i^2 \rangle\right]
\}^2\,, 
\end{equation}
is the iteration 0 ($k=0$), an initial guess of the parameters taken on the uncorrected variances. Summations in equations~(\ref{lsqFit1}) and (\ref{lsqFit2}) start at mode 5 in order to exclude tip-tilt and focus modes, because in practice, at the telescope, they include large contributions that do not come from the turbulence, such as wind-shake and vibrations. The logarithmic transformations in equations (\ref{lsqFit1}) and (\ref{lsqFit2}) are used to achieve a least squares model with homogeneous residuals.

Although it is not demonstrated that the problem is convex, we show by simulation that, for the cases considered, the algorithm converges. Non-convergence was never observed.

\subsection{Simulation setup}\label{sec:simulation}

The simulations to test the algorithm proposed in the previous section were conducted with the OOMAO toolbox  \citep{Conan2014}. 

In order to isolate   cross coupling from other known effects, atmospheric turbulence and SH-WFS measurements were treated in idealised conditions: turbulence modelled by  independent phase screens to assure good statistics, the sensor described by a noiseless geometrical model to avoid issues related with measurement errors \citep[e.g.][]{Thomas2006, Anugu2018} and noise artificially introduced directly on the modal variances. 

Sets of 5000 independent phase screens, $8\,\mathrm{m} \times 8\,\mathrm{m}$ in size, were generated for typical turbulence conditions  characterised by $\mathcal{L}_0= 32,  16, 8$ and 4 m and $r_0=10$ cm.  Only one $r_0$ is used because Zernike variances can be scaled to other values. Each set was sampled by a telescope with diameter $D=8\,\mathrm{m}$ without central obstruction and a SH-WFS (geometrical model) with $14 \times 14$ sub-apertures. A minimum light ratio condition of 75\% settled the number of valid sub-apertures at 148. The phases were reconstructed from the resulting open loop slopes using reconstruction matrices  $\mathbf{H}^+_r$  of size $(J(r)-1 \times 2N)$, with $J(r)$ representing the number of Noll modes up to radial  order $r$,  and $N$ the number of valid sub-apertures. Radial orders from 7 to 12 were considered, as these are typically referred in the literature \citep[e.g.] []{Schoeck2003,Fusco2004,Jolissaint2018}. All reconstruction matrices were obtained by singular value decomposition of sub-matrices obtained by truncation of a large (up to radial order $r=40$) matrix $\mathbf{G}$, constructed by collecting as columns the slopes of each individual Zernike mode.

The effects of noise were simulated by adding to the reconstructed noiseless Zernike coefficient  variances a term computed with equation~(\ref{noiseterm}), with $\sigma_0^2$ chosen to give specific values of signal-to-noise ratio (SNR) calculated by
\begin{equation}\label{SNR}
\mathrm{SNR}(r) = \frac{\sum_{j=5}^{J(r)}\langle b_j^2 \rangle}{\sum_{j=5}^{J(r)} \sigma^2_{\mathrm{n},j}}\,.
\end{equation}

Finally, the algorithm described in Section~\ref{algorithm} to estimate the turbulence parameters was tested on Zernike coefficient variances, obtained with varying reconstruction matrix sizes and number of iterations.

\section{Results and discussion}

In this section we start by addressing the effects of aliasing and cross coupling on the reconstructed Zernike coefficients variances.  Then the estimation of the turbulence parameters from these variances following the approach in Section~\ref{algorithm} is presented. The section ends with a discussion on the limitations and applicability of the algorithm to real world experiments. 

\subsection{Aliasing and cross coupling}\label{sec:results_aliasing}

\begin{figure}
\begin{center}

\pgfplotsset{ small, }

\begin{tabular}{l}
	\begin{tikzpicture}[baseline]
		\begin{semilogyaxis}[scale=1,
        xtick={ 15,78,136,190,253},
        xticklabels={ 15,78,136,190,253},
        xlabel={Noll modes},
		ylabel={Zernike coeff. variances / rad$^2$},
        legend entries={phase screens,reconstructed},legend style={font=\small,draw=none}]
		\addplot[mark=.] file {figures/data/data_px448_1008_ai2_ro21.dat};
		\addplot[color=red,mark=none] file {figures/data/data_px448_1008_bi2_ro21.dat};
        \node at (180,0) {21 radial orders};
		\end{semilogyaxis}
	\end{tikzpicture}
	\\
	\begin{tikzpicture}[baseline]
	\begin{semilogyaxis}[
        scale=1,
        xtick={ 15,78,136,210,276},
        xticklabels={ 15,78,136,210,276},
        xlabel={Noll modes},
		ylabel={Zernike coeff. variances / rad$^2$},
        ylabel near ticks,
    legend entries={phase screens,reconstructed},legend style={font=\small,draw=none}]
		\addplot[mark=.] file {figures/data/data_px448_1008_ai2_ro22.dat};
		\addplot[color=red,mark=none] file {figures/data/data_px448_1008_bi2_ro22.dat};
        \node at (180,0) {22 radial orders};
	\end{semilogyaxis}
	\end{tikzpicture}
\\
	\begin{tikzpicture}[baseline]
	\begin{axis}[
        ytick={190,210,231,253,276,300,325,351,378},
        yticklabels={189,,230,,275,,324,,377},
        xtick={190,210,231,253,276,300,325,351,378},
        xticklabels={18,,20,,22,,24,,26},
        scale=1,
        restrict x to domain=190:378,
        xlabel={$r$ (radial order)},
        ylabel={number of modes},
        legend entries={number of columns,rank of $\mathbf{H}_{\varparallel r}$},legend style={font=\small,draw=none,
        legend pos=north west}]
		\addplot[dotted] file {figures/data/aliasing_modes.dat};
		\addplot[color=red,mark=*,mark size=1pt] file {figures/data/aliasing_rankOfG.dat};
	\end{axis}
	\end{tikzpicture}
    
\end{tabular}	

\end{center}
\caption{
Onset of aliasing effects on the Zernike coefficient variances of reconstructed phases.
Top and middle: phase screens constructed with a limited number of modes from their 
original Zernike decomposition (21 and 22 radial orders, respectively). The phase reconstructions 
use reconstructor matrices with the same modes used in the phase screens generation in order 
to avoid cross-coupling. Aliasing, affecting the modal reconstructed variances, starts with
reconstructors containing 275 Noll modes - the number of modes (piston excluded) up to radial 
order 22. Bottom:  Rank of matrices $\mathbf{H}_{\varparallel r}$. Rank deficiency starts at 
radial order 22.
}\label{aliasing}
\end{figure}
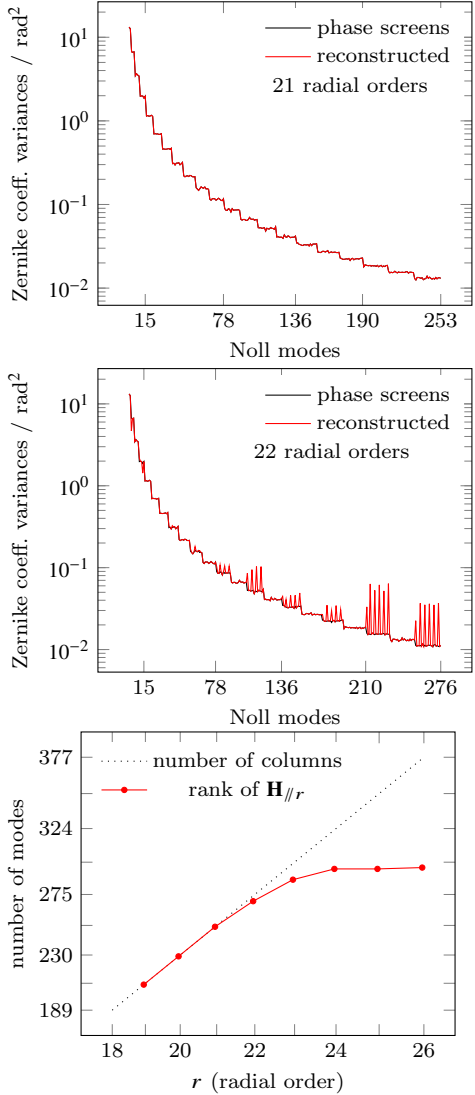

In order to illustrate the difference between aliasing and cross coupling effects, simulations with filtered phase screens were performed. These phase screens were constructed with a limited number of modes from the decomposition of the original ones and then sampled by the SH-WFS. Reconstruction with matrices containing the same modes eliminates perpendicular modes and thus, cross coupling. Since the SH-WFS measures the phase gradients with a finite resolution, as the order of the Zernike polynomials increases aliasing is expected at some point. Figure~\ref{aliasing} shows the results for phase screens and reconstruction matrices containing modes from 21 and 22 radial orders and the matrices rank as a function of their size.  The effects visible in the reconstructed modal variances for the 22 radial order case can be ascribed to aliasing. The rank of matrices $\mathbf{H}_{\varparallel r}$ becomes smaller than the number of columns  for $r \ge 22$, an indication that the columns produced by Zernike modes belonging to radial orders 22 and higher are linear combinations of previous columns. This is expected for the $14 \times 14$ SH-WFS under consideration, whose  Nyquist frequency is $f=7/D$. Following \cite{Conan1995} we have that $f\sim 0.3(r+1)/D$, which translates the Nyquist frequency into a maximum radial order $r \sim 22$.

\begin{figure}
\begin{center}

\pgfplotsset{
	small,
}
\begin{tabular}{l}
	\begin{tikzpicture}[baseline]
		\begin{semilogyaxis}[scale=1,
        xtick={4,22,37,56,79},
		ylabel={Zernike coeff. variances / rad$^2$},
        xticklabels={4,22,37,56,79},
    	xlabel={Noll modes},
        legend entries={phase screens ($r=12$),reconstructed ($r=12$)},legend style={font=\small,draw=none}]
		\addplot[mark=none] file {figures/data/data_px448_1008_ai2_ro12.dat};
		\addplot[color=red,mark=*,mark size=0.3pt] file {figures/data/data_px448_1008_bi2_ro12.dat};
		\end{semilogyaxis}
	\end{tikzpicture}
	\\
	\begin{tikzpicture}[baseline]
	\begin{semilogyaxis}[scale=1,
        legend entries={phase screens ($r=12$),reconstructed ($r=11$)},legend style={font=\small,draw=none},
		ylabel={Zernike coeff. variances / rad$^2$},
        xtick={4,22,37,56,79},
        xticklabels={4,22,37,56,79},
    	xlabel={Noll modes}]
		\addplot[mark=none] file {figures/data/data_px448_1008_ai2_ro12.dat};
		\addplot[color=red,mark=*,mark size=0.3pt] file {figures/data/data_px448_1008_bi2_ro12m1.dat};
	\end{semilogyaxis}
	\end{tikzpicture}
	\\
	\begin{tikzpicture}[baseline]
	\begin{semilogyaxis}[scale=1,
        legend entries={phase screens ($r=12$),reconstructed ($r=10$)},legend style={font=\small,draw=none},
		ylabel={Zernike coeff. variances / rad$^2$},
        xtick={4,22,37,56,79},
        xticklabels={4,22,37,56,79},
    	xlabel={Noll modes}]
		\addplot[mark=none] file {figures/data/data_px448_1008_ai2_ro12.dat};
		\addplot[color=red,mark=*,mark size=0.3pt] file {figures/data/data_px448_1008_bi2_ro12m2.dat};
	\end{semilogyaxis}
	\end{tikzpicture}
\end{tabular}%
\end{center}

\caption{
Effects of cross-coupling on the Zernike coefficient variances of reconstructed phases.
Variances from phase screens constructed using only 12 radial orders from their Zernike decomposition.
Reconstructed phase variances obtained from the slopes produced by these phase screens by using  reconstruction matrices with different number of modes: the same number of modes present in the  phase screens (top), one radial order less (middle) and two radial orders less (bottom).
}\label{crosscoupling}
\end{figure}
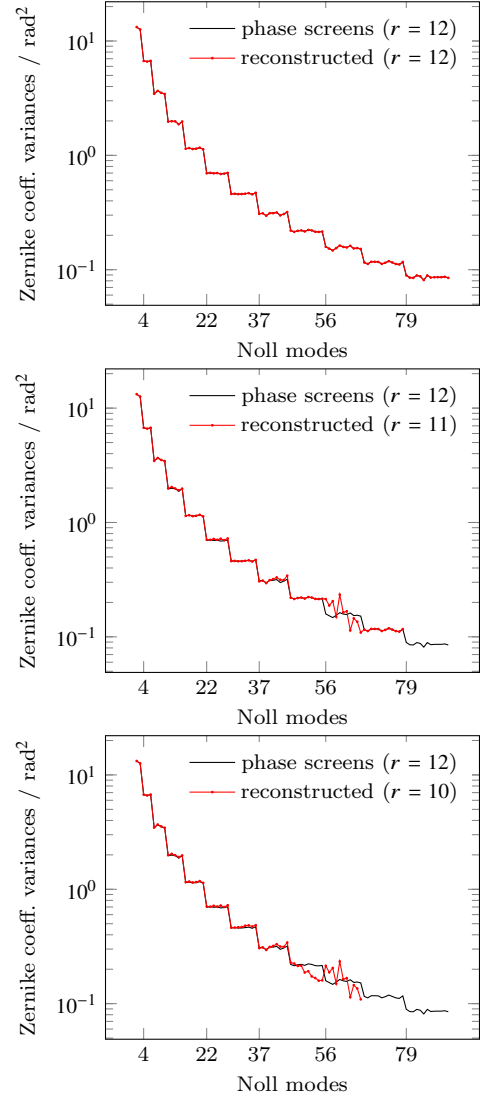

Cross coupling effects on the reconstructed modal variances are shown in Fig.~\ref{crosscoupling}. Filtered phase screens, this time containing 12 radial orders\footnote{The number 12 is somewhat arbitrary but representative of the maximum radial order used in AO SH-WFS atmospheric turbulence parameter estimation.}, were sampled by the SH-WFS and reconstructed with matrices containing 12, 11 and 10 radial orders. The existence of perpendicular modes in the last two cases leads to cross coupling.

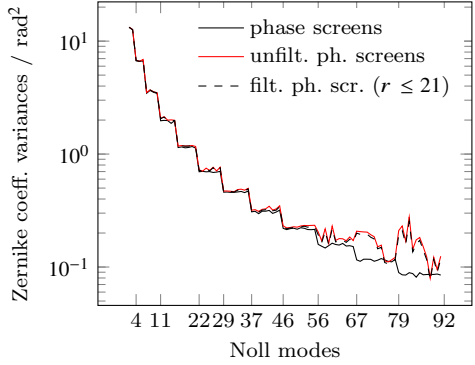
\begin{figure}
\begin{center}

\pgfplotsset{ small, }

	\begin{tikzpicture}[baseline]
		\begin{semilogyaxis}[scale=1,
        xtick={4,11,22,29,37,46,56,67,79,92},
        xticklabels={4,11,22,29,37,46,56,67,79,92},
		ylabel={Zernike coeff. variances / rad$^2$},
        xlabel={Noll modes},
        legend entries={
        phase screens,unfilt. ph. screens,filt. ph. scr. ($r\leq 21$)},
        	legend style={font=\small,draw=none},legend cell align=left]
		\addplot[mark=none] table[x index=0, y index=1] {figures/data/data_px448_aliasingvscc_ro12.dat};
		\addplot[color=red] table[x index=0, y index=2] {figures/data/data_px448_aliasingvscc_ro12.dat};
		\addplot[dashed]  table[x index=0, y index=3] {figures/data/data_px448_aliasingvscc_ro12.dat};
		\end{semilogyaxis}
	\end{tikzpicture}

\end{center}
\caption{Modal variances of phase screens and of phases reconstructed from unfiltered phase screens and from filtered phase screens (containing modes only up to radial order 21).
}\label{aliasingvscc}
\end{figure}

Unfiltered phase screens introduce, both aliasing and cross coupling effects. Modes above radial order 21 are  perceived as lower order modes and perpendicular modes contribute to the estimation of the the parallel ones (cf. equation~\ref{reconstructed_modes}).  Figure~\ref{aliasingvscc} displays the comparison between modal variances of phases reconstructed from slopes produced by unfiltered phase screens and by filtered phase screens with 21 radial orders (which do not introduce aliasing). Firstly, both reconstructed variances almost agree. Secondly, they both differ from the phase screens variances. This can be explained by a small contribution of aliasing from modes of radial order above 21 and a dominant contribution of cross coupling from modes of radial order below 22. It is concluded that cross coupling effects dominate over aliasing at the radial orders ($r\leq 12$) normally used for atmospheric turbulence estimation.

\subsection{Estimation of turbulence and noise parameters}

\begin{figure}

\begin{center}
\pgfplotsset{ small, }
        \begin{tikzpicture}[baseline]
                \begin{semilogyaxis}[scale=1,
        xtick={4,22,37,56,79},
                ylabel={Zernike coeff. variances / rad$^2$},
        xticklabels={4,22,37,56,79},
        xlabel={Noll modes},%
        legend entries={phase screens,reconstructed ($k=0$), theor. von Kármán.},legend style={font=\small,draw=none}]
                \addplot[only marks,mark=*,mark size = 0.6pt] table[x index=0, y index=1] {figures/data/data_estimation.dat};
                \addplot[only marks,color=red,mark=*,mark size=0.6pt] table[x index=0, y index=2] {figures/data/data_estimation.dat};
                \addplot[color=blue,mark=none] table[x index=0, y index=4] {figures/data/data_estimation.dat};
                \end{semilogyaxis}
        \end{tikzpicture}

\caption{Zernike coefficient variances, of modes from radial orders 2 (without focus) to 11, from simulations with the set of phase screens generated for 
$r_0=10\,\mathrm{cm}$ and $\mathcal{L}_0=8\,\mathrm{m}$.
In black, modal variances of the phase screens.  In red, modal variances of the reconstructed phases with $\mathrm{SNR}(r=9)=10$. 
In blue (solid line), von Kármán Zernike coefficient variances calculated with the turbulence parameters estimates obtained by applying algorithm of Section \ref{algorithm} to the reconstructed variances after three iterations ($k=3$).
}\label{estimation}

\end{center}
\end{figure}
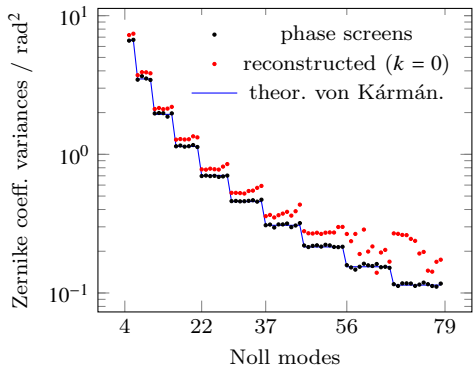

As presented in Section~\ref{algorithm}, the method estimates three parameters ($r_0$, $\mathcal{L}_0$ and $\sigma_0^2$) by fitting the reconstructed Zernike coefficient   variances, corrected, iteratively, for modal cross-coupling effects. An example of its application provides the results presented in Table \ref{estimation_tbl}, obtained from simulations with the set of phase screens generated for $r_0=10$ cm and $\mathcal{L}_0=8$ m (cf. Section \ref{sec:simulation}). 
Their modal variances are displayed in Fig.~\ref{estimation}, along with those of the respective reconstructions from the SH-WFS measurements. The latter exhibit disturbances due to noise and modal cross-coupling. Noise is expected by the fitting function~(\ref{fitfunction}) but not cross coupling. A fit to these reconstructed variances corresponds to iteration $k=0$ of the algorithm (\sout{Eq.} equation~(\ref{lsqFit2})) and produces biased estimates of the parameters.
At higher iterations, corrections for cross-coupling are calculated, using the turbulence parameters estimated at the previous iteration, and applied to the modal variances before the fit. At iteration $k=3$, the bias in the turbulence parameters is removed. The agreement is illustrated in Fig 4 where the von Kármán Zernike coefficient variances, calculated with the estimated turbulence parameters, are plotted against the modal variances of the phase screens. The retrieved noise parameter at $k=3$ compares well with the synthetic noise of $1.852\,\mathrm{rad}^2$ (corresponding to an $\mathrm{SNR}(r=9)=10$). This is not the case at $k=0$.

\begin{table} 
\begin{center} 
\caption{Turbulence and noise parameters estimates from modal variances of noiseless phase screens and noisy reconstructed phases without ($k=0$) and with ($k=3$) cross-coupling corrections. The  phase screens have $r_0=10\,\mathrm{cm}$ and $\mathcal{L}_0=8\,\mathrm{m}$.}\label{estimation_tbl}
\begin{tabular}{lccc} 
\hline
   & $r_0/\mathrm{cm} $ & $\mathcal{L}_0/\mathrm{m}$ & $\sigma^2_0/\mathrm{rad}^2$ \\ 
\hline
phase screens         & $9.99 \pm 0.05$  & $8.0 \pm 0.1$  &  $0.01 \pm 0.05$\\ 
reconstructed ($k=0$) & $11.4 \pm 0.4$   & $8.9 \pm 0.5$  &  $4.0  \pm 0.3$  \\
reconstructed ($k=3$) & $10.03 \pm 0.06$ & $8.0 \pm 0.1$  &  $1.95 \pm 0.06$\\ 
\hline
\end{tabular} 
\end{center} 
\end{table}

We now address the results of the turbulence parameters and noise estimation as a function of the maximum radial order used in the reconstruction and the number of algorithm iterations. The parameter values used in the generation of synthetic measurements are referred  in Section~\ref{sec:simulation}. Four noise regimes with SNR values of $\infty$, 1000, 100 and  10, were considered. Each SNR is calculated by \sout{Eq.} equation~(\ref{SNR}) at radial order 9 and is obtained by adjusting the value of the noise parameter $\sigma_0^2$. All fits are performed on sets of modal variances with lower radial order 2 from which the focus is excluded.

Figures~\ref{fig_r010_SNR10_r0} to \ref{fig_r010_SNR10_s2_pc} show,  for all considered outer scale values, the behaviour of the estimations as a function of  the reconstruction matrix size (indexed by the maximum radial order $r$) and the number of performed iterations ($k$).  For reference, the same estimations on the phase screens (ideal values) are also depicted (horizontal dotted line).

\input{figures/fig_r010_SNR10_r0.tex}
\input{figures/fig_r010_SNR10_L0.tex}
\input{figures/fig_r010_SNR10_s2_pc.tex}

\subsubsection{$k=0$}\label{sec:k_0}

We will focus, in this Section, on the $k=0$ regime. This is the standard approach,  apart from the denoising, without the iterative correction for cross coupling proposed in this article.

The impact of cross coupling is shown in Figures~\ref{fig_r010_SNR10_r0} to \ref{fig_r010_SNR10_s2_pc}, which display the estimates on noisy variances ($\mathrm{SNR}(r=9)=10$).

The estimates of $r_0$ and $\mathcal{L}_0$ have large uncertainties and, generally, large deviations (overestimations) from the expected values.  As the size of the reconstruction matrix ($r$) changes, different modes are affected by cross coupling and the distortions introduced in the Zernike variances (cf. Fig.~\ref{aliasingvscc}) change its position in  the spectrum leading to different estimates of $\mathcal{L}_0$ and $r_0$.  It should be underlined that the relative error does not scale monotonously with the  number of radial orders ($r$) used in the fit. This is a fundamental limitation of the standard method, where the size $r$ of the reconstruction matrix must be carefully chosen, to avoid large errors. The size of the reconstructor matrix which minimizes the  error due to cross coupling, $r\simeq 9$ in this case, depends on the geometry of the wavefront sensor and on its  modelling. For example, setting a different minimum light ratio condition may change this best size.

For the outer scale estimations depicted in Fig.~\ref{fig_r010_SNR10_L0} the fit quality decreases with increasing phase screen $\mathcal{L}_0$, as a consequence of poorer sampling of a large outer scale by the comparatively small aperture. The great sensitivity of the outer scale parameter to tip-tilt modes, leads, when excluding them from the fits, to occasional failures in the estimations (nonphysical large values) at large outer scales and for some sizes of the reconstructor matrix.  When observed, these difficulties cease with the inclusion of the tip-tilt modes in the fits.

With regards to the noise parameter, it is not well estimated by the standard method ($k=0$). It is underestimated for maximum radial order $r<10$ and overestimated for $r\ge 10$.

\subsubsection{$k>0$}\label{sec:k+}

The case $k>0$ is now addressed. This is the iterative approach proposed in this article.

Figures~\ref{fig_r010_SNR10_r0} to \ref{fig_r010_SNR10_s2_pc} show that, regardless of the reconstruction matrix size, $r$, at iteration 3, all $r_0$ and
$\mathcal{L}_0$ estimations have converged to values in agreement, within the estimation uncertainty, with the expected ones (direct estimations on the phase screens). The  same is true for the noise parameter convergence to the reference values (except for  some cases involving larger reconstructor matrices, where nevertheless the difference is very small).

The better the estimations at $k=0$, the faster the convergence is achieved. For $r=9$, $r_0$ and $\mathcal{L}_0$ estimations have converged, or are close to converge, as soon as iteration 1.

An important aspect that emerges from the results is the robustness of the proposed iterative algorithm. Significant improvements are achieved even when the 
corrections are calculated using inaccurate parameter values from estimations at the previous iteration. This is particularly evident in the $\mathcal{L}_0$ estimations. Extreme examples are those for which the zero iteration gives non-physically  large outer scale values  (in the km range)  and, nevertheless, the corrections computed with these values allow reasonable estimations at the first iteration.

\subsubsection{Summary of results}\label{sec:summary}

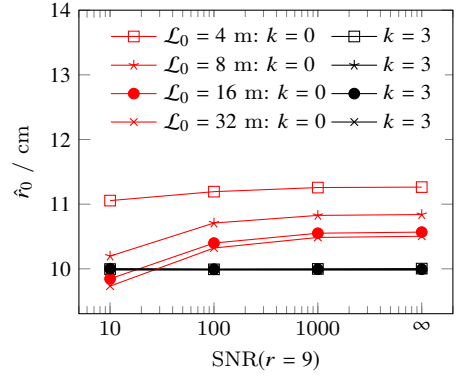
\begin{figure}

\begin{center}
\pgfplotsset{ small }


\begin{tikzpicture}[baseline,trim axis left]
	\begin{semilogxaxis}[
    scale=1.0, 
    xtick={1,10,100,1000,10000}, 
    xlabel={SNR$(r=9)$},
	xticklabels={1,10,100,1000,$\infty$},
    ymax=14,
	ylabel={$\hat{r}_0$ / cm},
    legend entries={ 
    				 $\mathcal{L}_0=4$ m: $k=0$,
                     $k=3$, 
    				 $\mathcal{L}_0=8$ m: $k=0$,
                     $k=3$, 
    				 $\mathcal{L}_0=16$ m: $k=0$,
                     $k=3$, 
    				 $\mathcal{L}_0=32$ m: $k=0$,
                     $k=3$, 
                     },
    legend style={font=\small,draw=none},
    legend cell align=left,
    legend columns=2
        ]
\pgfplotstableread{

10.0000   11.0552    9.9964   10.1968    9.9929    9.8460   10.0021    9.7329    9.9820
100.0000   11.1935    9.9924   10.7083    9.9911   10.3987    9.9946   10.3222    9.9812
1000.0000   11.2569    9.9985   10.8264    9.9911   10.5508    9.9936   10.4853    9.9810
10000.0000  11.2641    10.0037  10.8396    9.9918   10.5680    9.9935   10.5037    9.9810

}\repchundred 
\addplot[mark=square,red]    table[x index=0,y index=1]
{\repchundred};
\addplot[mark=square]    table[x index=0,y index=2]
{\repchundred};
\addplot[mark=star,red]    table[x index=0,y index=3]
{\repchundred};
\addplot[mark=star] table[x index=0,y index=4] 
{\repchundred};
\addplot[mark=*,red] table[x index=0,y index=5] 
{\repchundred};
\addplot[mark=*] table[x index=0,y index=6] 
{\repchundred};
\addplot[mark=x,red] table[x index=0,y index=7] 
{\repchundred};
\addplot[mark=x] table[x index=0,y index=8] 
{\repchundred};
	\end{semilogxaxis}
\end{tikzpicture}

\end{center}

\caption{Fried parameter estimations as a function of $\mathrm{SNR}(r=9)$, for  reconstructions with $r=9$ and $r_0=10\,\mathrm{cm}$. Note that the $k=3$ curves overlap.}\label{fig_r0_vs_snr}
\end{figure}
\begin{figure}

\begin{center}
\pgfplotsset{ small }

\begin{tikzpicture}[baseline,trim axis left]
	\begin{loglogaxis}[
    scale=1.0, 
    xtick={1,10,100,1000,10000}, 
    xlabel={SNR$(r=9)$},
	xticklabels={1,10,100,1000,$\infty$},
    ymax=1024,
	ylabel={$\hat{\mathcal{L}}_0$ / m},
    ytick={4,8,16,32}, 
	yticklabels={4,8,16,32},
    legend entries={ 
    				 $\mathcal{L}_0=4$ m: $k=0$,
                     $k=3$, 
    				 $\mathcal{L}_0=8$ m: $k=0$,
                     $k=3$, 
    				 $\mathcal{L}_0=16$ m: $k=0$,
                     $k=3$, 
    				 $\mathcal{L}_0=32$ m: $k=0$,
                     $k=3$, 
                     },
    legend style={font=\small,draw=none},
    legend cell align=left,
    legend columns=2
        ]
\pgfplotstableread{
    10.0000    4.4852    4.0018    8.6645    7.9732   16.5996   15.8576   31.9717   31.8866
    100.0000    4.4861    4.0002    9.0711    7.9689   18.9303   15.8303   47.0856   31.8405
    1000.0000    4.4878    4.0006    9.1664    7.9686   19.6558   15.8267   54.1671   31.8328
    10000.0000	 4.4880	   4.0006    9.1771    7.9691   19.7412   15.8253   55.1104   31.8319

}\repchundred 
\addplot[mark=square,red]    table[x index=0,y index=1]
{\repchundred};
\addplot[mark=square]    table[x index=0,y index=2]
{\repchundred};
\addplot[mark=star,red]    table[x index=0,y index=3]
{\repchundred};
\addplot[mark=star] table[x index=0,y index=4] 
{\repchundred};
\addplot[mark=*,red] table[x index=0,y index=5] 
{\repchundred};
\addplot[mark=*] table[x index=0,y index=6] 
{\repchundred};
\addplot[mark=x,red] table[x index=0,y index=7] 
{\repchundred};
\addplot[mark=x] table[x index=0,y index=8] 
{\repchundred};
	\end{loglogaxis}
\end{tikzpicture}

\end{center}

\caption{Outer scale estimations as a function of $\mathrm{SNR}(r=9)$ for reconstructions with $r=9$ and $r_0=10$ cm.}
\label{fig_L0_vs_snr}
\end{figure}
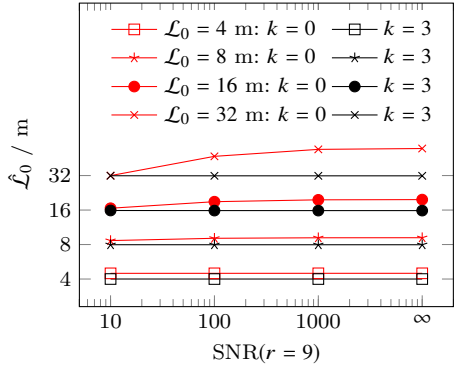

In this section, we summarize the top-level results, in terms of $r_0$, $\mathcal{L}_0$ and signal-to-noise, obtained with the conventional approach ($k=0$) and with the proposed algorithm after 3 iterations ($k=3$), for the case of fits on nine radial orders of modes ($r = 9$). 

Figure~\ref{fig_r0_vs_snr} shows that the Fried parameter value estimated without iteration is biased by the signal-to-noise ratio, being over-estimated most of the time. The over-estimation increases with signal-to-noise ratio. The bias is stronger for smaller outer scales. In contrast, the iterative method is robust to this bias, with all curves overlapping at the expected value. 

With regards the outer scale, it is also biased by the signal-to-noise ratio, but with a smaller amplitude (cf. Fig.~\ref{fig_L0_vs_snr}) and mostly for larger outer scales.

This apparently counter-intuitive behaviour, more noise - better initial ($k=0$) estimations, stems from the fact that cross coupling effects arise in very localised regions of the Zernike spectrum and thus, bias the estimation to higher values. On the other hand noise, which has a more uniform distribution across all radial orders, masks the cross coupling disturbances.

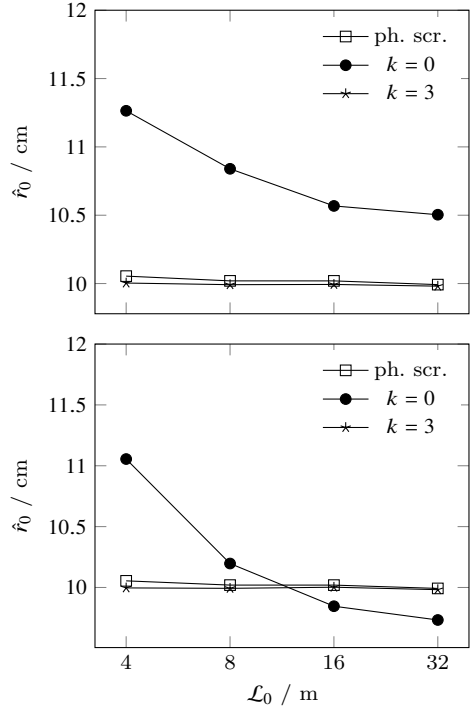
\begin{figure}

\begin{center}
\pgfplotsset{ small }

\begin{tabular}{rl}

\begin{tikzpicture}[baseline,trim axis left]
	\begin{semilogxaxis}[
    scale=1.0, 
    xtick={4,8,16,32}, 
	xticklabels={,,},
    ymax=12,
	ylabel={$\hat{r}_0$ / cm},
    legend entries={ ph. scr. ,  $k = 0$ , $k = 3$},
    legend style={font=\small,draw=none}
        ]
\pgfplotstableread{

     4.0000   10.0544   11.2641   10.0037   20.1763   22.5197   20.1485 
     8.0000   10.0196   10.8396    9.9918   20.0581   21.7120   20.0370
    16.0000   10.0195   10.5680    9.9935   20.0329   21.2508   20.0083
    32.0000    9.9922   10.5037    9.9810   19.9888   21.0240   19.9575

}\repchundred 
\addplot[mark=square]    table[x index=0,y index=1] {\repchundred};
\addplot[mark=*]    table[x index=0,y index=2] {\repchundred};
\addplot[mark=star]    table[x index=0,y index=3] {\repchundred};
	\end{semilogxaxis}
\end{tikzpicture}
\\

\begin{tikzpicture}[baseline,trim axis left]
	\begin{semilogxaxis}[
    scale=1.0, 
    xtick={4,8,16,32}, 
    xlabel={$\mathcal{L}_0$ / m},
	xticklabels={4,8,16,32},
    ymax=12,
	ylabel={$\hat{r}_0$ / cm},
    legend entries={ ph. scr. ,  $k = 0$ , $k = 3$},
    legend style={font=\small,draw=none}
        ]
\pgfplotstableread{

    4.0000   10.0544   11.0552    9.9964   20.1763   22.2737   20.1513
    8.0000   10.0196   10.1968    9.9929   20.0581   20.4384   20.0444
   16.0000   10.0195    9.8460   10.0021   20.0329   19.7068   20.0228
   32.0000    9.9922    9.7329    9.9820   19.9888   19.4569   19.9564

}\repchundred 
\addplot[mark=square]    table[x index=0,y index=1] {\repchundred};
\addplot[mark=*]    table[x index=0,y index=2] {\repchundred};
\addplot[mark=star]    table[x index=0,y index=3] {\repchundred};
	\end{semilogxaxis}
\end{tikzpicture}

\end{tabular}
\end{center}

\caption{Fried parameter estimations as a function of $\mathcal{L}_0$ for simulations with $r_0=10$\,cm and $r=9$. Top - no noise, bottom - $\mathrm{SNR}(r=9)=10$.
}\label{fig_r0_vs_L0}
\end{figure}

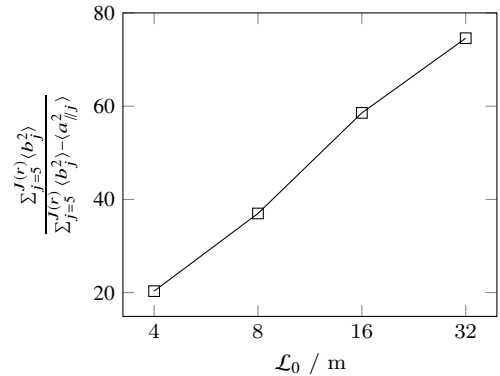
\begin{figure}

\begin{center}
\pgfplotsset{ small }

\begin{tikzpicture}[baseline,trim axis left]
	\begin{semilogxaxis}[
    scale=1.0, 
    xtick={4,8,16,32}, 
    xlabel={$\mathcal{L}_0$ / m},
	xticklabels={4,8,16,32},
	ylabel={ $\frac{\sum_{j=5}^{J(r)}\langle b_j^2 \rangle}
             {\sum_{j=5}^{J(r)} {\langle b_j^2 \rangle - 
             \langle a_{\varparallel j}^2 \rangle }}$ }
        ]
\pgfplotstableread{

     4.0000   20.32   20.16    
     8.0000   36.98   37.13   
    16.0000   58.56   59.61   
    32.0000   74.58   72.92   

}\repchundred 
\addplot[mark=square]    table[x index=0,y index=1]
{\repchundred};
	\end{semilogxaxis}
\end{tikzpicture}

\end{center}

\caption{Signal to cross coupling effects ratio at $r=9$ as a function of
$\mathcal{L}_0$, for $r_0=10\,\mathrm{cm}$.}\label{fig_stcc_vs_L0}
\end{figure}

\begin{figure}

\begin{center}
\pgfplotsset{ small }

\begin{tabular}{rl}

\begin{tikzpicture}[baseline,trim axis left]
	\begin{loglogaxis}[
    scale=1.0, 
    ytick={4,8,16,32,64}, 
    xtick={4,8,16,32,60,120,240,480}, 
    yticklabels={4,8,16,32,64}, 
	xticklabels={,,},
    ymax=150,
	ylabel={$\hat{\mathcal{L}}_0$ / m},
    legend entries={ ph. scr. ,  $k = 0$ , $k = 3$},
    legend style={font=\small,draw=none},
    legend cell align=left,
    legend pos=north west 
        ]
\pgfplotstableread{
    4.0000    4.0123    4.4880    4.0006   60.0000    3.9980    4.4879    3.9992
    8.0000    7.9786    9.1771    7.9691  120.0000    8.0292    9.2378    8.0208
   16.0000   15.9160   19.7412   15.8263  240.0000   16.1073   20.3344   16.0533
   32.0000   32.0918   55.1104   31.8319  480.0000   31.7353   54.3504   31.4354

}\repchundred 
\addplot[mark=square]    table[x index=0,y index=1] {\repchundred};
\addplot[mark=*]    table[x index=0,y index=2] {\repchundred};
\addplot[mark=star]    table[x index=0,y index=3] {\repchundred};
	\end{loglogaxis}
\end{tikzpicture}
\\

\begin{tikzpicture}[baseline,trim axis left]
	\begin{loglogaxis}[
    scale=1.0, 
    xtick={4,8,16,32,60,120,240,480}, 
    ytick={4,8,16,32,64}, 
    yticklabels={4,8,16,32,64}, 
    xlabel={$\mathcal{L}_0$ / m},
	xticklabels={4,8,16,32,4,8,16,32},
    ymax=150,
    xmin=0,
	ylabel={$\hat{\mathcal{L}}_0$ / m},
    legend entries={ ph. scr. ,  $k = 0$ , $k = 3$},
    legend style={font=\small,draw=none},
    legend cell align=left,
    legend pos=north west 
        ]
\pgfplotstableread{
    4.0000    4.0123    4.4852    4.0018   60.0000    3.9980    4.4861    3.9995
    8.0000    7.9786    8.6645    7.9732  120.0000    8.0292    8.7176    8.0217
   16.0000   15.9160   16.5996   15.8576  240.0000   16.1073   16.8227   16.0758
   32.0000   32.0918   31.9717   31.8866  480.0000   31.7353   31.5055   31.4098

}\repchundred 
\addplot[mark=square]    table[x index=0,y index=1] {\repchundred};
\addplot[mark=*]    table[x index=0,y index=2] {\repchundred};
\addplot[mark=star]    table[x index=0,y index=3] {\repchundred};
	\end{loglogaxis}
\end{tikzpicture}

\end{tabular}
\end{center}

\caption{Outer scale estimations as a function of $\mathcal{L}_0$
for simulations with $r_0=10$ cm and $r=9$. Top - no noise, bottom - $\mathrm{SNR}(r=9)=10$.
}\label{fig_L0_vs_L0}
\end{figure}
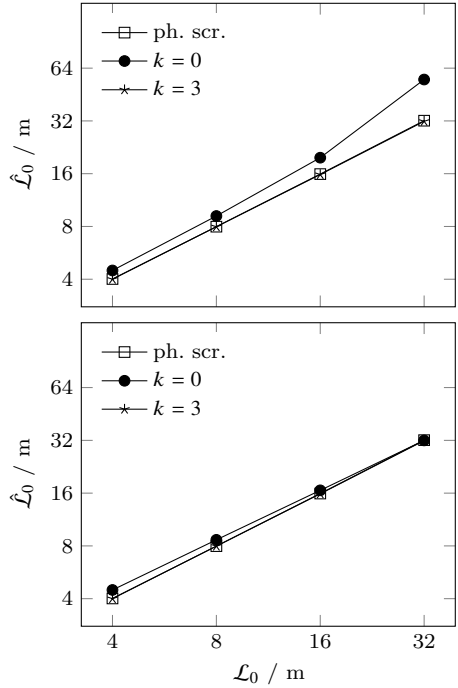

Figure~\ref{fig_r0_vs_L0}, presents the bias in the estimated Fried parameter versus the outer scale in the phase screens. It  clearly illustrates that the estimated Fried parameter is also biased by the outer scale. As the outer scale increases, the $r_0$ estimations improve in precision and accuracy. This is because the ratio of the Zernike coefficient variances to the disturbances produced by cross coupling, increases as the outer scale increases, as shown in Fig.~\ref{fig_stcc_vs_L0}. 

With regards to the outer scale $\mathcal{L}_0$ estimations depicted in Fig.~\ref{fig_L0_vs_L0}, it is clear that up to $\mathcal{L}_0 = 32\,\mathrm{m}=4D$ the fit converges to the phase screen value.

\begin{table} 
\begin{center} 
\caption{Percent error (bias) in estimations of $r_0$ after 3 iterations for $r = 9$,  $r_0=10\,\mathrm{cm}$. \label{tab:r0_r9_bias} }
\begin{tabular}{ccccc} 
\hline \diagbox{SNR}{$\mathcal{L}_0/\mathrm{m}$} & 4 & 8 & 16 & 32 \\ 
\hline
$\infty$  & -0.50  & -0.28  & -0.26  & -0.11\\
 1000  & -0.56  & -0.28 &  -0.26 &  -0.11\\
 100  & -0.62  & -0.28  & -0.25  & -0.11\\
 10         & -0.58  & -0.27  & -0.17  & -0.10\\
\hline
\end{tabular} 
\end{center} 
\end{table}

\begin{table} 
\begin{center} 
\caption{Percent error (bias) in estimations of $\mathcal{L}_0$ after 3 iterations for $r = 9$,  $r_0=10\,\mathrm{cm}$.  \label{tab:L0_r9_bias} }
\begin{tabular}{ccccc} 
\hline
  \diagbox{SNR}{$\mathcal{L}_0/\mathrm{m}$} & 4 & 8 & 16 & 32 \\ 
\hline
$\infty$  & -0.29 & -0.12  & -0.56  & -0.81\\ 
1000     & -0.29 & -0.13  & -0.56  & -0.81\\
100     & -0.30 & -0.12  & -0.54  & -0.78\\ 
10     &  0.26 & -0.07  & -0.37  & -0.64\\ 
\hline
\end{tabular} 
\end{center} 
\end{table}

In Tables~\ref{tab:r0_r9_bias} and \ref{tab:L0_r9_bias} the bias for the proposed method, in \%, of the estimated parameters, with regards to the phase screens values is presented for the range of signal-to-noise ratios considered. It is of less than $1\%$ for the range of simulation parameters considered. The non-dependence of the bias on the signal-to-noise ratio is due to the inclusion in the iterative algorithm of the variance noise as a parameter.

\subsection{Limitations and applicability to experiments}

The proposed algorithm for correcting cross coupling relies on the correct modelling of the atmospheric turbulence and the SH-WFS measurements.  In order to isolate   cross coupling from other known effects, these two aspects were  treated in idealised conditions (cf. Section~\ref{sec:simulation}): turbulence modelled by independent phase screens, the sensor is described by a noiseless geometrical model and the noise is artificially introduced directly on the modal variances. 

In real applications, departures from the above simplified framework are expected. In the present work open loop telemetry is used, but in an on-sky application it is desirable to use closed loop telemetry, without disturbing the scientific observation. Synthetic open loop telemetry must be generated from closed loop telemetry. A simplified version of the algorithm is already in place at ESO's AO Facility  to provide turbulence parameters estimations for all of its laser guide star wavefront sensors, estimations that have shown to be very consistent from one WFS to the other, and with external seeing monitors as well. An accurate estimation with this improved algorithm of the seeing in the line of sight of the science instruments would greatly benefit the quality classification of the observations, their scheduling and the understanding of the delivered AO performance. 

Another aspect of real applications is the signal-to-noise ratio. Following \cite{Rigaut1992} the $\mathrm{SNR}(r=9) = 10$ in this work translates in a natural guide star magnitude of  $m_\mathrm{V}=12$ for a NAOS AO-like system. When applied to natural guide star SH-WFSs present in many instruments, it would provide an unprecedented accuracy and reliability for noise-limited measurements. On the other hand, the experience of the AO Facility shows that the laser guide star measurements are not signal-to-noise limited \citep{Kolb2017}.

\section{Conclusions}

Motivated by the ubiquity of AO SH-WFS and it's advantages in atmospheric turbulence parameters estimation the effects of aliasing and cross coupling in the estimation are analysed for the first time for single sensor SH-WFS AO telemetry.

For the adopted SH-WFS, with $14 \times 14$ lenslets, and in the range of reconstructed modes considered (7 to 12 radial orders), the Zernike coefficient variances were found to be affected  essentially by modal cross coupling, responsible for distortions which, depending on the  reconstructor size, can lead to large inaccuracies in the estimation of the turbulence parameters.

A method which removes from the modal variances the disturbances created by cross coupling is proposed. The   theoretical Zernike coefficient variances fitting function is adapted with an added term modelling measurement noise. The variances to be fitted are corrected for  cross coupling. Since these corrections require the knowledge of the turbulence parameters, the algorithm is iterative.

Simulated measurements of the SH-WFS, obtained at different noise regimes (guide star of magnitude 12 or brighter) and atmospheric conditions (outer scales from 4\,m to 32\,m at Fried parameter of 10\,cm) were used to test the algorithm. The results showed the elimination of the bias to sub-percent level in the estimation of the turbulence parameters after three or less iterations, regardless of the number of reconstructed modes. 

\section*{Acknowledgements}

The research leading to these results has received partial funding from the European Union’s Horizon 2020 research and innovation programme under Grant Agreement 730890 (OPTICON) and the Portuguese Fundação para a Ciência e a Tecnologia with grant number UID/FIS/00099/2013. CC was supported by the A*MIDEX project (no. ANR-11-IDEX-0001-02) funded by the ”Investissements d’Avenir” French Government programme, managed by the French National Research Agency (ANR). The authors thank the referee for the valuable comments that improved the article content.




\bibliographystyle{mnras}
\bibliography{cca} 


\bsp	
\label{lastpage}
\end{document}